\documentclass[graybox]{svmult}

\usepackage{mathptmx}       % selects Times Roman as basic font
\usepackage{helvet}         % selects Helvetica as sans-serif font
\usepackage{courier}        % selects Courier as typewriter font
\usepackage{type1cm}        % activate if the above 3 fonts are
\usepackage{makeidx}         % allows index generation
\usepackage{graphicx}        % standard LaTeX graphics tool
\usepackage{multicol}        % used for the two-column index
\usepackage[bottom]{footmisc}% places footnotes at page bottom

\def\pdot {\dot \textrm{P}}

\def\Omdot {\dot \Omega}
\def\ltsima{$\; \buildrel < \over \sim \;$}
\def\lsim{\lower.5ex\hbox{\ltsima}}
\def\gtsima{$\; \buildrel > \over \sim \;$}
\def\gsim{\lower.5ex\hbox{\gtsima}}

\def\uu {4U~0142$+$61}
\def\kes {1E~1841$-$045}

\def\rxs {1RXS~J170849.0$-$400910}
\def\ee {1E~2259$+$586}
\def\xte {XTE~J1810$-$197}
\def\cxo {CXOU~J164710.2$-$455216}
\def\smc {CXOU~J010043.1$-$721134}
\def\zeroq {SGR~0418$+$5729}
\def\zeroc {SGR~0501$+$4516}
\def\zerosei {SGR~1806$-$20}
\def\zerozero {SGR~1900$+$14}
\def\sedici {SGR~1627$-$41}
\def\lmc {SGR~0526$-$66}
\def\qui {1E 1547.0$-$5408}
\def\oo {1E~1048.1$-$5937}
\def\xdzeroquattro {RX~J0420.0$-$5022}
\def\xdzerosette {RX~J0720.4$-$3125}
\def\xdzerootto {RX~J0806.4$-$4123}
\def\xdtredici {RX~J1308.6$+$2127}
\def\xdsedici {RX~J1605.3$+$3249}
\def\xddiciotto {RX~J1856.5$-$3754}
\def\xdventuno {RX~J2143.0$+$0654}

\def\rat {PSR~J1819$-$1458}

\makeindex             % used for the subject index
\begin{document}

\title*{X-ray emission from isolated neutron stars}
% Use \titlerunning{Short Title} for an abbreviated version of
% your contribution title if the original one is too long
\author{Sandro Mereghetti}
% Use \authorrunning{Short Title} for an abbreviated version of
% your contribution title if the original one is too long
\institute{Sandro Mereghetti \at INAF, IASF-Milano, v. E.Bassini 15, I-20133 Milano, Italy \email{sandro@iasf-milano.inaf.it}
%\and Name of Second Author \at Name, Address of Institute \email{name@email.address}
}
%
% Use the package "url.sty" to avoid
% problems with special characters
% used in your e-mail or web address
%
\maketitle

%\abstract*{ }

\abstract{X-ray emission is a common feature of all varieties of isolated neutron stars (INS)
and, thanks to the advent of sensitive instruments with good spectroscopic, timing, and imaging
capabilities, X-ray observations have become an essential tool in the study of these objects.
Non-thermal X-rays from young, energetic radio pulsars have been detected since the beginning of
X-ray astronomy, and the long-sought thermal emission from cooling neutron star's surfaces
can now be studied in detail in many pulsars spanning different ages, magnetic fields, and, possibly,
surface compositions.
In addition, other different manifestations
of INS have been discovered with X-ray observations.
These new  classes of high-energy sources, comprising the nearby X-ray Dim Isolated Neutron Stars,
the Central Compact Objects in supernova remnants,
the Anomalous X-ray Pulsars, and the Soft Gamma-ray Repeaters, now add up to several tens of
confirmed members,  plus many candidates, and allow us to study  a
variety of phenomena unobservable in "standard'' radio pulsars.
}

\section{Introduction}
\label{sec:introduction}

With more than 1800 detections, rotation powered  pulsars (RPP) constitute
by far the largest class of isolated neutron stars (INS),
despite  only one out of $\sim$10 radio pulsars is visible because of beaming.
Accounting for the selection effects of radio observations,
a total population of $\sim10^6$  RPP is estimated for  the whole
Galaxy \cite{fau06}.
Observations at gamma-ray energy, where pulsar beaming angles are larger,
are now contributing to increase the number of known RPP \cite{abd09_blind}.
About one hundred  RPP have  been detected also at X-ray energies \cite{bec09}:
they include  the youngest and more energetic pulsars (like the Crab),
a few older neutron stars at  small distances (e.g. Vela and Geminga)
and several tens of recycled millisecond pulsars (most of which are found in globular clusters \cite{gri09}).

All kinds of INS, not only the RPP, are X-ray emitters.
X-ray observations have been crucial to discover other manifestations of INS,
that for various reasons were missed in the standard searches for
radio pulsars.  The nearby X-ray Dim Isolated Neutron Stars (XDINS),
the Central Compact Objects (CCOs) in supernova remnants, the Anomalous
X-ray Pulsars (AXPs), and the Soft Gamma-ray Repeaters (SGRs) are examples of these
new classes, which, despite totalling  only a few tens of sources, are
particularly  interesting because they offer a different view on a
variety of phenomena unobservable in "standard'' radio pulsars.

\begin{figure}[b]
%\sidecaption
\includegraphics[scale=.37, angle=0]{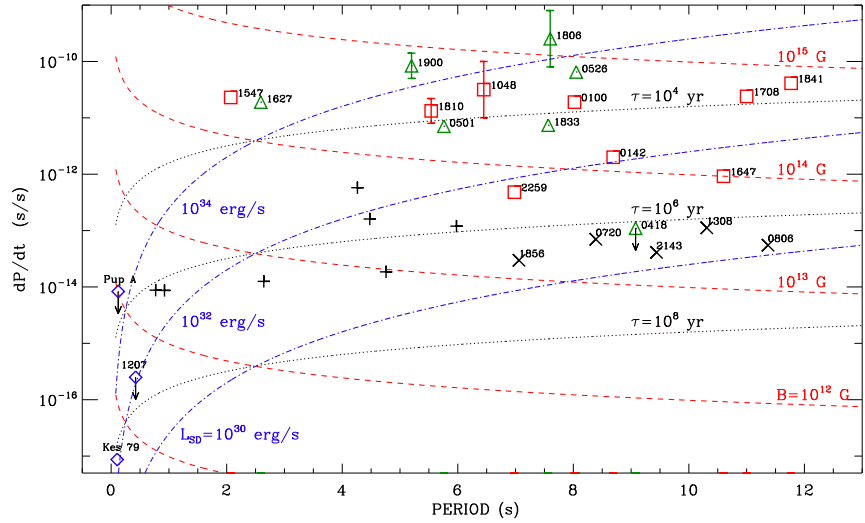}
\caption{P--$\pdot$ diagram for different classes of INS:
XDINS ($\times$), CCO ($\diamond$), AXP (squares), SGR ($\bigtriangleup$),
RRAT (+). The vertical bars indicate the range of $\pdot$ variations
observed in SGR and AXP.
Lines of constant magnetic field (dashed), characteristic age (dotted), and spin-down
luminosity (dash-dotted) are also indicated.}
\label{fig:ppdot}
\end{figure}

This review is focussed on the  main properties of the  X-ray emission from these
new classes of INS. For an excellent review of the X-ray emission from RPP see \cite{bec09}.
The location  on a
P--$\pdot$ plot diagram of some of the  objects discussed here is
shown in Fig. \ref{fig:ppdot}. The figure also gives  lines
of constant magnetic field (computed assuming dipole braking,
B = 3.2 10$^{19}$ (P $\pdot$)$^{1/2}$ G), characteristic age ($\tau_c$=$\frac{1}{2}$P/$\pdot$), and
spin-down luminosity (L$_{SD}$ = I $\Omega$ $\Omdot$ =
4 10$^{46}$ $\pdot$ / P$^3$ erg s$^{-1}$).

\section{Origin of the X-ray emission in isolated neutron stars}
\label{sec:ori}

The X-ray emission observed from INS can be powered by internal heat, rotational energy, accretion, and
magnetic field decay. The relative importance of these energy sources,
that can also operate at the same time, depends on the age and  physical properties of the neutron star.

Neutron stars have internal temperatures of $\sim10^{11}$ K at birth,
that rapidly drop to $\sim10^{9}$ K.
For the following $\sim10^5-10^6$ years  the dominant cooling mechanism
is neutrino emission from the star's isothermal core.
This leads to surface temperatures of several $10^5$ to $10^6$ K, with  thermal emission peaking in the soft X-ray band.
Temperature gradients on the star's surface generally produce observable modulations at
the rotation period. Thermal X-ray emission
can be observed in INS with ages of $\sim10^4-10^6$ years,
provided they are sufficiently close and not too absorbed by the interstellar medium.
Older neutron stars are too cool to significantly emit  X-rays,
while in the youngest pulsars the thermal radiation is difficult to detect
because it is outshined  by the brighter non-thermal emission. A recent review on the thermal emission
from neutron star is given by \cite{zav09}.

Non-thermal emission, that extends over a broad energy range, originates from
charged particles accelerated in the NS magnetospheres at the expense
of rotational energy (see, e.g., \cite{che09}).
Non-thermal  X-rays are characterized by power-law spectra and  strongly
anisotropic emission patterns, giving rise to large pulsed fractions.
The pulse profiles often show narrow (double) peaks, but in many cases nearly
sinusoidal profiles are observed.
The most luminous RPP are the Crab and two young pulsars in the
Large Magellanic Cloud
(the only three pulsars with L$_{SD}>10^{38}$ erg s$^{-1}$).
The efficiency with which the rotational energy is converted to non-thermal luminosity
is about 10$^{-3}$ \cite{bec97,li08}, but there is a
large dispersion around the average value \cite{pos02}, as expected because of
different viewing orientations and, possibly,  also other effects.
Studies of the L$_X$--L$_{SD}$ relation with a large sample are complicated  by the
fact that a significant fraction of the rotational energy
loss powers pulsar wind nebulae,
which are difficult to disentangle in the more distant and/or fainter objects
without adequate spatial resolution.

Accretion is a well established process in X-ray binaries (e.g., \cite{fra02}).
In the lack of a companion star,
accreting matter could originate directly from the interstellar medium.
However, the relatively large space velocity of neutron stars, and the low density of the
interstellar medium, make this process unable to provide sufficiently high luminosities.
Alternatively, the  matter could  be supplied by a debris disk
formed by fall-back in the supernova explosion that produced the neutron star.
Although this possibility  seems more promising,
no unambiguous  evidence for an INS powered by accretion has been found yet.

The relevance of magnetic energy in powering the emission from  neutron stars has
been recognized only recently, with the observation of SGRs and
AXPs (see \cite{woo06, mer08} for reviews). High magnetic fields (B$\sim$10$^{14}$--10$^{15}$ G)
were first invoked to explain the SGRs and, in particular, the unique
properties of the exceptional event of March 5, 1979 from \lmc\ \cite{pac92, dun92}.
Objects in which magnetic field decay is the
dominant energy source have been named Magnetars. They are particularly interesting
because they offer the unique possibility to study physical processes
in magnetic fields of unequalled strength.

%%%%%%%%%%%%%%%%%%%%%%%%%%%%%%%%%%%%%%%%%%%%%%%%%%%%%%%%%%%%%%55
\begin{table}
\caption{Isolated Neutron Stars}
\label{tab:list}
\begin{tabular}{lccccl}
\hline\noalign{\smallskip}
Name  & P & $\pdot$       & D$^{(a)}$              &  Association & Notes$^{(b)}$  and \\
      &  (s)  &  (s s$^{-1}$)  & (kpc) &   &    References  \\
\noalign{\smallskip}\svhline\noalign{\smallskip}
%\hline
 \multicolumn{6}{c}{XDINS} \\
 \hline
\xdzeroquattro\ & 3.45  &  -    &  0.345   & - &  \cite{hab04b} \\[3pt]
\xdzerosette\ & 8.39  & 7.01 10$^{-14}$ &  0.36    & - &   G? \cite{hab97,kap07,van07,hoh10} \\[3pt]
\xdzerootto\ & 11.37 & (5.5$\pm$3.0) 10$^{-14}$  &  0.25    & - &  \cite{hab04b,kap09b} \\[3pt]
\xdtredici\  & 10.31 & 1.120 10$^{-13}$  & 0.5     & - & RBS1223 \cite{ham02,kap05a} \\[3pt]
\xdsedici\  &   -   &   -     & 0.39   & - & RBS1556 \cite{mot99,zan06} \\[3pt]
\xddiciotto\  & 7.06 & (2.97$\pm$0.07) 10$^{-14}$ & 0.16   & - &  \cite{wal01,tie07,van07a,van08} \\[3pt]
\xdventuno\  &  9.43 & (4.1$\pm$1.8) 10$^{-14}$  & 0.43 & - & RBS1774 \cite{zam01,kap09d} \\[3pt]
2XMM J104608.7--594306    &  -     &  -      &  2   & - & candidate \cite{pir09} \\[3pt]
RX J1412.9+7922 &  -    &  -       &  3.6 &   - & candidate, Calvera \cite{rut08,she09}\\[3pt]
\hline
 \multicolumn{6}{c}{CCO} \\
 \hline
RX J0822.0--4300 & 0.122 & $<$8 10$^{-15}$& 2.2 & Puppis A & \cite{pet96,got09} \\[3pt]
CXOU J085201.4--461753 & -      & -       &   1  & G266.1--1.2  & \cite{pav01} \\[3pt]
1E 1207.4--5209 & 0.424 & $<$2.5 10$^{-16}$& 2 & G296.5+10.0 &  \cite{mer02,del04,got04} \\[3pt]
CXOU J160103.1--513353 &  -  &  - & 5  & G330.2+0.1 &  \cite{par06} \\[3pt]
RX J1713.4--3949  & - & - & 1.3 & G347.3--0.5 & \cite{laz03} \\[3pt]
CXOU J185238.6+004020  & 0.105 & 8.7 10$^{-18}$ & 7.1 & Kes 79 & \cite{hal07,hal10} \\[3pt]
1E 161348--5055.1  & -  & -  & 3.3  & RCW 103 &   T \cite{del06}\\[3pt]
CXOU J232327.8+584842  &  - &  - &  3.4 & Cas A & \cite{cha01,mer02c,pav09} \\[3pt]
XMMU J172054.5--372652  & - & - &   4.5 &  G350.1--0.3 & candidate \cite{gae08} \\[3pt]
XMMU J173203.3--344518  & - & - &   3.2 & G353.6--0.7 & candidate \cite{tia10}\\[3pt]
CXOU J181852.0--150213  & - & - &  8.5 & G15.9+0.2 & candidate \cite{rey06b} \\[3pt]
\hline
 \multicolumn{6}{c}{AXP and SGR} \\
 \hline
\smc  & 8.02 & 1.9 10$^{-11}$ &60 & SMC &  AXP \cite{lam02,mcg05,tie08} \\[3pt]%\cite{lam02,mcg05,tie08}  \\[3pt]
\uu  & 8.69 & 2 10$^{-12}$&  3.6     &  - &    B, G? \cite{isr94,dib07b} \\[3pt]
\oo & 6.45 & (1--10) 10$^{-11}$ &  8 & - &   B, G \cite{sew86,tie05a,gav02,dib09} \\[3pt]
\qui  & 2.07 & 2.3 10$^{-11}$ &5 & G327.2--0.1 &   T, B, R \cite{cam07c,hal08,esp08b,mer09}     \\[3pt]
PSR J1622--4950 & 4.33 & 1.7 10$^{-11}$ & 9 & - &   R  \cite{lev10} \\[3pt]
\cxo  & 10.6 & 9.2 10$^{-13}$ & 3.9 & Westerlund 1 &    T, B, G  \cite{mun07,isr07b} \\[3pt]
\rxs  & 11.0 & 2.4 10$^{-11}$  & 5& - &   G \cite{sug97,isr99,dal03,dib07a} \\[3pt]
\xte & 5.54 &(0.8--2.2) 10$^{-11}$&3.1 & - &   T, B, R \cite{ibr04,cam06,woo05,ber09} \\[3pt]
\kes & 11.77& 4.1 10$^{-11}$ & 8.5 & Kes 73 &   G \cite{vas97,got02,mor03,dib07a} \\[3pt]
\ee     & 6.98 & 4.8 10$^{-13}$&4 &CTB 109&    B, G \cite{fah81,kas03} \\[3pt]
AX J1844.8--0256 & 6.97 & - & 8.5 & G29.6+0.1 & candidate, T \cite{vas00,tam06}\\[3pt]
\hline
\zeroq  &  9.1 & $<$1.1 10$^{-13}$     &   2   & -  &   T, B \cite{van10,esp10} \\[3pt]
\zeroc  &  5.76 & 7.1 10$^{-12}$   & 1.5 &   - &   T, B \cite{rea09,apt09,eno09} \\[3pt]
\lmc  & 8.05 & 6.5 10$^{-11}$ & 55 & LMC, N49&  B, GF \cite{maz79,maz99b,tie09} \\[3pt]
\sedici & 2.59 & 1.9 10$^{-11}$  &    11 & - &   T, B \cite{woo99c,mer06a,esp09a,esp09b}      \\[3pt]
\zerosei&7.6&(8--80) 10$^{-11}$&8.7&star cluster&  B, GF  \cite{kou98,pal05,mer05c,woo07} \\[3pt]
SGR 1833--0832 &  7.6  & 7.4 10$^{-12}$  &   10    & -  &   T, B \cite{gog10,esp10b} \\[3pt]
\zerozero&5.2 &(5--14) 10$^{-11}$ &15&star cluster &  B, GF, G? \cite{maz79b,woo99b,mer06b}\\[3pt]%
\hline
\noalign{\smallskip}\hline\noalign{\smallskip}
\end{tabular}

$^{(a)}$ in several cases the distances have large uncertainties; the values
adopted in the figures of this work are indicated here.

$^{(b)}$ B = bursts, G = glitches, GF = giant flares, R = radio emission, T = transient

\end{table}
%%%%%%%%%%%%%%%%%%%%%%%%%%%%%%%%%%%%%%%%%%%%%%%%%%%%%%%%%%%%%%%%%%%%%%%%%%%%%%%%%%%%%%%%%%%

\section{The X-ray Dim Isolated Neutron Stars}
\label{sec:xdins}

A handful of X-ray sources with
large X-ray-to-optical flux ratios,  F$_{x}$/F$_{opt}\sim$10$^4$--10$^5$,
typical of INS, were discovered in the ROSAT satellite All Sky Survey.
Their INS nature was confirmed by the measurement of a large proper
motion in the brightest source, \xddiciotto\ \cite{wal01}, and by the discovery
of pulsations at few seconds in other sources \cite{hab97,ham02}.
The original members of this class\footnote{they have been nicknamed ''The Magnificent Seven'' (M7, hereinafter)} and two new candidates
are listed in Table \ref{tab:list}. Recent reviews on the XDINS are given in  \cite{hab07,tur09}.

XDINS have very soft thermal spectra (blackbody temperatures T$_{BB}\sim$40--110 eV), X--ray luminosities
L$_X\sim10^{30}-10^{32}$ erg s$^{-1}$, spin periods in the 3--12 s range,
faint optical counterparts (V$>$25), and no radio emission.
Period derivatives of the
order of 10$^{-14}$--10$^{-13}$ s s$^{-1}$ have been   measured for several XDINS through
phase connected timing, but in some cases these values are   still
poorly constrained   (see Table \ref{tab:list}).
The  temperatures of XDINS
are consistent with neutron stars cooling curves if they have ages
of 10$^5$--10$^6$ years and  the effect of their
strong magnetic field  is taken into account \cite{agu08}.
It is thus generally believed that the XDINS  are powered by residual thermal energy.

X-ray sources with such soft spectra can only observed if the interstellar
absorption is small:  all the M7 are closer than $\sim$0.5  kpc and
have  N$_H<$4 10$^{20}$ cm$^{-2}$.
Their proximity allowed in several cases the  measurement of parallax and
and proper motion
showing a distribution of  transverse velocities  consistent
with that of radio pulsars \cite{mot09}.
For these velocities, and typical ISM densities, accretion from the ISM, originally invoked
to explain the XDINS emission, cannot  provide the observed luminosity.
Although the timing parameters give spin-down ages
$\tau_c$$\sim$ 1--2 million years,
the measured spatial velocities and likely birth places in nearby OB associations
strongly suggest that their true ages are smaller (see Table \ref{tab:ages}).

%%%%%%%%%%%%%%%%%%%%%%%%%%%%%%%%%%%%%%%%%%%%%%%%%%%%%%%%%%%%%%55
\begin{table}
\caption{Comparison of estimated and spin-down ages of INS} \label{tab:ages}
\begin{tabular}{lccccl}
\hline\noalign{\smallskip}
Source  & Spin-down   &  Estimated      &  Method  \\
      &  age ($\tau_c$ )  &      age &     \\
\noalign{\smallskip}\svhline\noalign{\smallskip}
\xdzerosette\  & 1.9 Myr  &  (0.7$^{+0.3}_{-0.2}$) Myr  & proper motion  \\[3pt] %\cite{kap07}
\xdtredici\  & 1.5 Myr  & (0.5--1.4) Myr & proper motion  \\[3pt] %\cite{mot09}
\xddiciotto\ & 3.8 Myr & 0.4 Myr  & proper motion \\[3pt] %\cite{van08,wal01}
\hline
\kes    &  4.5 kyr   &  0.5--1 kyr   & SNR age      \\[3pt] %\cite{tia08}
\ee     & 230 kyr & (11.7$\pm$1.2) kyr   & SNR age    \\[3pt] %\cite{sas04}
\lmc    &  2 kyr   &  5 kyr  & SNR  age  \\[3pt] % \cite{van92}
\hline
CXOU J085201.4--461753 & $>$240 kyr &  3.7 kyr &  SNR  age   \\[3pt]
1E 1207.4$-$5209 & $>$27 Myr & 7 kyr &   SNR  age \\[3pt]
CXOU J185238.6+004020  & 190  Myr  &  7 kyr & SNR  age  \\[3pt]
\hline
\noalign{\smallskip}\hline\noalign{\smallskip}
\end{tabular}

%$^{a}$ the two values refer to the non-coherent and coherent timing analysis, respectively.

\end{table}
%%%%%%%%%%%%%%%%%%%%%%%%%%%%%%%%%%%%%%%%%%%%%%%%%%%%%%%%%%%%%%%%%%%%%%%%%%%%%%%%%%%%%%%%%%%

Despite various attempts, none of the XDINS has been detected
in the radio band. The most recent  upper limits  of  0.14--5 $\mu$Jy kpc$^2$,
obtained at 1400 MHz \cite{kon09}, correspond to pulsed luminosities
well below those of the  faintest observed radio PSRs.
However, given the small sample and considering the anticorrelation
between beaming aperture angle and pulse period, it cannot be excluded that the lack of radio emission be
simply due to unfavorable orientations.

The X-ray light curves of XDINS are nearly sinusoidal, with  pulsed fractions,
ranging from 1.5\% to $\sim$20\%, and  moderate energy dependence. Only
in the  case of \xdtredici\ two peaks are clearly seen. These properties indicate that
we are likely seeing emission from a large fraction of the star's surface.

The first  spectra of XDINS, obtained with ROSAT, could be well described with blackbody curves.
With the better data of XMM and Chandra, significant deviations from this simple model have been detected in most XDINS (the only  exception is  \xddiciotto\ \cite{bur03}).
The features consist of broad absorption lines  in the energy range $\sim$0.2--0.8 keV
(see \cite{van07a,hab07} and references therein).
Multiple lines, with energy spacings  consistent with harmonic ratios,
are present in nearly half of the sources.
They have been interpreted as proton cyclotron lines or
as H and/or He atomic transitions, falling in the soft X-ray range
due to the effect of a high magnetic field on the binding energies.
Both interpretations require   B$\sim10^{13}-10^{14}$ G,
broadly consistent with the values of 2.5 and 3.4 10$^{13}$ G derived from the timing
parameters of  \xdzerosette\ and \xdtredici\ assuming  dipole
braking\footnote{Note that the lines measure the surface field, which can be higher
than the dipole component  dominating at large radii and responsible for the spin-down.}.

Variability on a time scale of few years was discovered in
\xdzerosette\ \cite{dev04}, a target regularly observed for XMM calibrations
because it was believed to be a constant source.
%Also the light curves changed.
If we exclude AXPs and SGRs, this has been the first INS  to show
significant changes in its X-ray flux, spectrum and pulse profile.
Based on the first observations,
it was suggested that the variations
had  a periodicity of 7.1$\pm$0.5 years  and could be caused by precession of the
neutron star \cite{hab06}.
However, the most recent analysis, which include new observations and
take into account the variable phase shifts between hard and soft energy bands, do not
support the periodicity  \cite{hoh09}.
Most of the spectral variation occurred in a relatively short timescale
of half a year \cite{van07}, in coincidence with a possible glitch.
A change in the temperature and/or composition of part of the NS surface might
have been caused by the energy released in the glitch, or alternatively
by a sudden accretion episode.

The other XDINS seem to be constant in flux, but small variations might have been unnoticed
since, with the exception of \xddiciotto\  ,  they have not been observed very often with sensitive instruments.
On the other hand, \xddiciotto\ is twice as bright as \xdzerosette\ and  has  been observed repeatedly,
but no variability has been reported. Interestingly this is the only XDINS without spectral lines
\cite{bur03} and with the smallest pulsed fraction ($\sim$1.5\% \cite{tie07}).

Until recently, attempts to enlarge the XDINS sample had little success.
Ongoing searches in the  ROSAT Bright Sources Catalogue are reported in
\cite{tur10}. Looking for new INS among the tens of thousands serendipitous
XMM and Chandra sources requires a long and difficult multi-wavelength effort,
also considering that they are fainter and more distant than the bright M7.
One promising source has been   selected
among XDINS candidates discovered with XMM \cite{pir09}, but its temperature
(kT$_{BB}$=117 eV) and  estimated luminosity ($\sim10^{33}$ erg s$^{-1}$) are larger
than those of the M7.
The same is true for another candidate, RX J1412.9+7922  \cite{rut08},
for which other interpretations  (e.g. a nearby millisecond pulsar) cannot be excluded yet.
Its blackbody temperature is kT$_{BB}\sim$200 eV. A   hydrogen
atmosphere model (kT$\sim$122 eV), with a possible emission line at 0.53 keV,
gives a better fit to the Chandra data \cite{she09}.
If the emission comes from the whole star's surface, its distance would be 3.6 kpc,
while a smaller emission region  and  distance  should give rise to
pulsations.  Only the detection of a periodicity could clarify the real
nature of these new XDINS candidates.

\section{ Central Compact Objects in Supernova Remnants}
\label{sec:cco}

The group of Central Compact Objects (CCOs) consists of several radio quiet
X--ray sources located at the center of shell-like
SNRs and with  high F$_x$/F$_{opt}$\cite{pav04,del08}.
The association  with SNRs (see Table \ref{tab:list})
imply ages at most of a few tens kyrs, and gives the possibility  to know their distances.
Three CCOs show X-ray pulsations with periods in the 0.1--0.4 s range,
consistent with  young neutron stars,
but with unexpectedly  small  spin-down rates  \cite{got07,hal10}.
The extremely small $\pdot$  measured for the  CCO in  Kes 79
implies a dipole magnetic field of only 3.1 10$^{10}$ G, while the
upper limits for the sources in G296.5+10.0 and Puppis A
give B$<$3.3  10$^{11}$ G and  B$<$10$^{12}$ G,   respectively.
The  spin-down ages of these three CCOs exceed by orders of magnitude the  ages
of their associated SNRs (see Table \ref{tab:ages}), indicating that these INS were probably
born with spin periods very close to the current values.
It is likely that their relatively long initial spin period and low
magnetic field be causally connected. This has led to the ''anti-magnetar''
nickname for these objects.
The three ''anti-magnetars'' show  a  variety of different  pulse profiles: the CCOs
in Kes 79 has a pulsed fraction of  \gtsima 60\%,  one of the highest among thermally
emitting INS \cite{hal07}. A more standard value of \ltsima10\%
is shown in 1E 1207--5209, but with the
peculiarity that most of the pulsation can be attributed to phase variations in the
shape of its prominent absorption lines at 0.7, 1.4, 2.1 and possibly 2.8 keV \cite{del04}.
The pulsations in the Puppis A CCOs eluded an earlier detection because
of a phase shift of 180$^{\circ}$ between nearly sinusoidal profiles in the
soft ($<$1.2 keV) and hard bands \cite{got09}.

Most CCOs have thermal-like X-ray emission, with  blackbody temperatures in the
range $\sim$0.2--0.5 keV,
and no evidence for additional non-thermal components.
The rotational energy loss of the three pulsed COOs is certainly too small to give
a detectable contribution to their observed X-rays, assuming a typical efficiency
of 10$^{-3}$. The emission could be due to residual cooling and/or weak accretion
from a residual disk.

When data of good statistical quality are available,
single-temperature blackbody  spectra do not provide acceptable fits.
Better fits are obtained with the sum of two blackbodies  with kT$_1$
in the range 0.16--0.4 keV and kT$_{2}\sim$2 kT$_{1}$.
This simple model can be thought as
a first approximation of a non-uniform temperature distribution. The hotter blackbody
is energetically significant:
the bolometric luminosities of the two components are in
the ratio  L$_2$/L$_1$$\sim$0.2--0.7.
The two-blackbody model implies small X-ray emitting areas (R$_1$$\sim$0.4--4 km,
R$_2$$\sim$0.06--0.8 km),
which are difficult to explain in weakly magnetized neutron stars.
The situation does not  significantly change if  atmosphere models, which
give lower temperatures and larger radii, are used instead of the blackbody fits.

It is unclear whether also the CCOs with unknown spin period
can  be interpreted as ''anti-magnetars''.
The lack of pulsar wind nebulae and of radio/gamma emission suggests that
also these  COOs have a small L$_{SD}$, but this can result equally well  from
a small $\pdot$ or from a long spin period.
Thus it cannot be excluded that CCOs constitute a
heterogenous class comprising objects of different kinds.
The source
at the center of the $\sim$330 years old  Cas A  SNR is one of the most studied CCOs.
It has a very soft  spectrum and   most spectral fits, with a variety of thermal models, indicate
an emitting area smaller than the whole surface of a neutron star.
No pulsations have been detected, down to quite stringent limits \cite{mer02c,pav09,hal10},
but an anti-magnetar interpretation remains plausible, considering the possibility of an unfavorable
orientation and the variety of pulsed fractions seen in other INS.

On the other hand,  the CCO  in RCW 103, with a clear periodicity
at 6.67 hours and long term variations spanning more than two orders of magnitude
\cite{del06}, has unique properties
requiring an ad-hoc explanation. The periodicity, if interpreted as an
orbital motion, suggests a low mass binary \cite{del06,bha09},
but the deep optical limits pose severe constraints on the companion star \cite{del08b}.
Models involving magnetars, isolated or in a synchronous binary, and
fall-back disks have also been proposed \cite{piz08,li07}.

\section{The Magnetar candidates: Anomalous X-ray Pulsars and Soft Gamma-ray Repeaters}
\label{sec:magnetars}

Anomalous X-ray Pulsars (AXPs) and  Soft Gamma-Ray Repeaters (SGRs) are
spinning-down pulsars characterized by a luminosity in the soft and hard X-rays
larger than their available rotational energy loss and by the emission of bursts
and flares.

The first AXPs, discovered as bright pulsars in the soft X-ray range ($<$10 keV),
were initially classified as X-ray binaries powered by accretion.
Further X--ray data, coupled to deep  searches for optical/IR
counterparts, revealed their different nature \cite{mer95}. In
particular, these observations showed that the narrow period
distribution, long-term spin-down, and soft spectrum of the AXPs
were at variance with the properties of the larger
population of pulsars in massive binaries. Furthermore, no signs of binary companions
could be found in the AXPs \cite{mer98}.

The SGRs were instead discovered through the detection of bright and short bursts in the hard X-ray/soft
gamma-ray range, and initially considered  a particular subclass of gamma-ray bursts
\cite{lar86,att87}, with the notable property of ``repeating''
from the same sky direction. When accurate localizations became
available, it was possible to identify the X-ray counterparts of
SGRs, finding that they are pulsating sources very similar to the AXPs.

Although historically divided in two classes, many similarities indicate
that AXPs and SGRs are probably the same kind of astrophysical  objects
(see \cite{woo06,kas07,mer08} for reviews).
The most successful model to explain AXPs and SGRs  involves
highly magnetized neutron stars, or ``magnetars'' \cite{dun92,
tho95, tho96}, but other possibilities have been
proposed, e.g., models based on INS accreting
from residual disks \cite{cha00a,per00b,alp01,ert09}, or different kinds of
quark stars \cite{xu07,hor07,ouy07a,cea06}.

Magnetars differ from the normal pulsars, not only
for their higher field intensity, but also because their magnetosphere is not
dipolar. It is believed to consist of a twisted dipole, i.e. a field with  a significant
azimuthal component. This causes the presence of large-scale currents
with high charge density flowing in the magnetosphere and
affecting the emerging spectra by resonant scattering.

At energy below 10 keV, the X-ray spectra of AXP and SGRs have been traditionally described with
a two-components model consisting of a power-law plus a blackbody with kT$_{BB}\sim$0.5 keV.
The steepness of the power-law component (especially in the AXPs,
photon index $\sim$3--4) requires  N$_H$ values higher
than those independently estimated in other ways and leads to
overestimate the flux of the near infrared and optical counterparts (unless a
drastic, and possibly un-physical, spectral cut-off is invoked).
This problem is solved by adopting a two-blackbody model, that
gives equally good fits \cite{hal05}. XMM observations of \smc\ in  the Small
Magellanic Cloud, less affected by the interstellar absorption than the
Galactic AXPs/SGRs, have clearly demonstrated  that the power-law plus
blackbody model can be rejected with high confidence, while a good
fit to the soft X-ray spectrum is obtained with the sum of two blackbodies \cite{tie08}.

Magnetars have  been detected also in the hard X-ray range (\gtsima 20 keV), with INTEGRAL and RXTE.
Hard power-law tails, extending to $\sim$150 keV, have been seen both in AXPs \cite{kui04,den08a,den08b}
and SGRs \cite{mer05a,goe06b,esp07,rea09}.
The hard X-ray emission is pulsed, and gives a luminosity  comparable to, and in some cases higher than,
that in the soft X-ray band.

Considerable effort is ongoing in the analysis and interpretation of the AXPs and SGRs spectra,
in order to progress from purely phenomenological fits  to more physical models.
The presence of a relatively dense plasma in magnetospheres with a twisted configuration \cite{tho02}
is expected to affect the emergent spectrum through resonant cyclotron
scattering  (RCS) of the thermal photons emitted by the underlying
neutron star's surface. A simplified model, based on  a
semi-analytical treatment of these effects \cite{lyu06b}, has been
systematically applied to several magnetar candidates
\cite{rea08}. This showed that at E$<$10 keV the RCS model can replace
the blackbody plus power-law model (in most sources an additional
power-law is still required to fit the hard X-ray tails). More
realistic 3-D simulations have also been performed to compute the
magnetar's spectral models \cite{fer07,nob08a,nob08b} and quite
successfully applied to the spectra of several magnetar candidates
\cite{zan09}.

\begin{figure}[b]
%\sidecaption
\includegraphics[scale=.4, angle=-90]{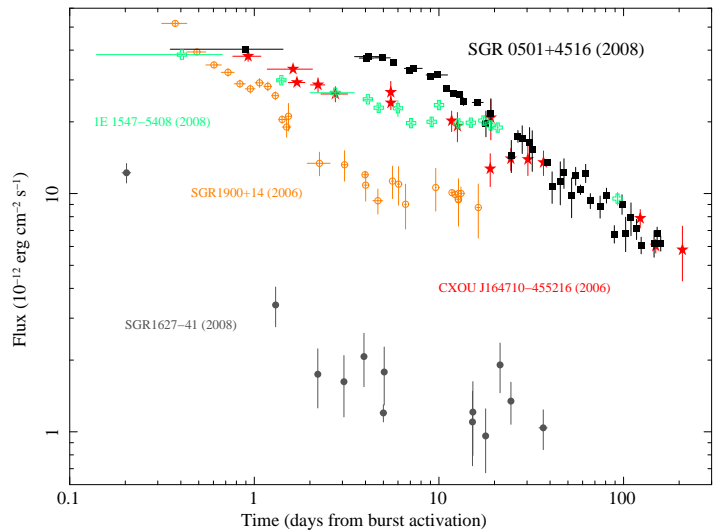}
\caption{X-ray light curves of  outburst decays observed
in several AXPs and SGRs (from \cite{rea09}).}
\label{fig:dec}
\end{figure}

Two possibilities have been proposed to explain the
high-energy tails in the context of the twisted
magnetosphere model  \cite{tho05}:
bremsstrahlung from a thin turbulent layer of
the star's surface heated to kT$\sim$100 keV by magnetospheric
currents and synchrotron emission from mildly relativistic pairs
produced at a height of $\sim100$ km above the neutron star.
Compton up-scattering of soft thermal photons, emitted from the
star surface, by relativistic electrons threaded in the
magnetosphere has  also been
considered \cite{har91,gon00,bar07}.

Compared to all the other INS, AXPs and SGRs stand apart for their
striking variability properties.
I refer not only to the  short bursts and  (giant) flares, but
also to the so called ''persistent'' emission,
which for historically bright sources varies around  10$^{36}$ erg s$^{-1}$,
while for newly-discovered transients has been seen to span up to five
orders of magnitude.
About  half of the AXPs/SGRs have been always seen at relatively high
flux levels since the time of their discovery.
They were thought to be steady sources, but, as more sensitive imaging observations
became available, it appeared that variations  of $\sim$50\%
on timescales from days to months are quite common.
Long term variations are sometimes correlated with changes in the spin-down properties.
This is not unexpected in the magnetar model involving variations in the magnetospheric
twist angle.
For example, the luminosity variations, the correlation between spectral hardness and spin-down
rate, and the increase in the bursting activity observed before the 2004
giant flare of \zerosei\ \cite{mer05c}
are  consistent with a gradually increasing magnetospheric twist angle.
This gives rise to a larger optical depth for resonant scattering (causing a
hardening of the X-ray spectrum),  a  higher rate of crustal fractures
(responsible for the bursts), and  a faster spin-down
(because of the larger fraction of field lines that open out across
the speed of light cylinder).
The spectral softening and the decrease in the flux
and spin-down rate observed after the giant
flare \cite{rea05b,tie05b} are also consistent with this picture, suggesting a
twist angle reduction as consequence of the global magnetic rearrangement
following the giant flare.

The first transient AXP, \xte ,  was serendipitously discovered during the decay phase
of an outburst in 2003 \cite{ibr04}. Comparison of the highest observed luminosity
with archival data, showing a  faint soft source with L$_{x}\sim10^{33}$ erg s$^{-1}$,
indicated a dynamic range of two orders of magnitude.
Several other transient AXP/SGR have been discovered after \xte\ (see Table \ref{tab:list}).
When bright, their spectral and timing properties are similar to those of the
persistent sources.
During their low (or ''quiescent'') states  they have luminosity
of $\sim10^{32}$ erg s$^{-1}$ and soft thermal spectra, that make them similar to the CCOs.

Transient AXPs/SGRs now outnumber the persistent members of the class, and their
number will certainly increase in the future.
It is  difficult to estimate how many AXP/SGRs are present
in the Galaxy, because the duty cycle of transient magnetars is poorly known.
At least two sources have shown more than one outburst
(\sedici\ in 1998 and in 2008 \cite{esp08};
\qui\ in Summer 2007 \cite{hal08} and 2008--2009 \cite{mer09}),  but other events might have been missed.
In fact the rate of discoveries of new transients has increased in the latest years,
thanks to the presence of satellites with large field of view and continuous sky monitoring,
like Fermi and Swift.  These new data revealed a variety of behaviors for what concerns
the outburst evolution of the different sources, as shown by the examples plotted in Fig. \ref{fig:dec}.
Poorly sampled light curves can be roughly described by power-law decays  with a variety of slopes,
while the well sampled  light curves show that the time decay is initially steeper and
flattens after about one day. This could indicate the presence of two different mechanisms operating
on different timescales, e.g. dissipation of magnetospheric currents followed by cooling of the star's
crust. Re-brightenings, or ''bumps'', superimposed on the decaying trends have also been observed, for
example in \zerozero\ \cite{fer03}.

The observation of outburst decays, and, more in general, of the spectral and flux variability of transient
AXPs/SGRs, has the potential to greatly advance our understanding of neutron stars.
For example,  detailed studies of the long term spectral evolution have been reported
for \xte\ \cite{got05,ber09}.   However some
caveats should be remembered when interpreting the observational data.
Often the observations have a poor and irregular time sampling, especially in the initial and possibly
more variable phases, and/or are obtained with different instruments.
This reduces the possibility of determining correlations between the different energy
ranges in a robust way.
At late times, when the fluxes are small, spectral uncertainties can be significant,
also in view of the bolometric corrections required by the  high interstellar absorption.
Furthermore, the observed fluxes most likely result from the superposition of
various physical components with different variability behaviors and difficult to disentangle.
Part of these problems will be reduced when broad-band, sensitive instruments
able to react in a short time to the discovery of new transient events will become available.

\section{Rotating Radio Transients}
\label{sec:rrat}

New techniques of analysis for transient radio signals, applied to data
of the Parkes Multibeam Survey,  have led to the discovery
of  Rotating Radio Transients (RRATs) \cite{mcl06}:  neutron stars which emit
short (2-30 ms) radio pulses at intervals of minutes
to hours. Their rotation periods, ranging from  0.4 to 7 s, have been inferred from the largest common divisors of the time intervals between bursts. RRATs are probably a large galactic
population, that remained undiscovered for a long time due to lack of adequate radio searches,
but their relation to other classes of INS is unclear (see \cite{mcl09b} for a review).

The radio properties of RRATs show analogies with those observed in some peculiar
radio RPP, and they might represent extreme cases of some of these phenomena.
On the other hand, their long rotation periods, and the fact that transient
radio emission has been observed in a couple of AXPs \cite{cam06,cam07c}, suggest a possible
relation with magnetars.
The  period derivatives, determined to date for only $\sim$6--7 of the nearly
twenty known RRATs \cite{mcl09}, indicate  magnetic fields above 10$^{13}$ G
only in about half of the cases (see Fig. \ref{fig:ppdot}).

\rat\ is the only RRAT that has been detected at X-ray energies \cite{rey06}.
It has a luminosity of $\sim$(2--5) 10$^{33}$ erg s$^{-1}$  (for d=3.6 kpc) and its
X-ray emission is pulsed at the rotation period of 4.26 s. The   light curve is nearly
sinusoidal, with a pulsed fraction of about 35\%. The spectrum,   well fit by a
blackbody (T$_{BB}\sim$0.14 keV) plus  a possible feature around 1 keV
\cite{mcl07},  indicates that the emission is of thermal origin (the luminosity
is subject to the distance uncertainty, but most likely larger than L$_{SD}$=3 10$^{32}$ erg s$^{-1}$).
A Chandra observation has shown evidence for diffuse X-ray emission surrounding \rat\  \cite{rea09b}.
If this nebula is powered by the rotational energy of the RRAT, an
efficiency  of $\sim$20\%, much higher than that of all the other pulsar wind nebulae, is required.
This discrepancy could be reduced by a smaller distance, or by invoking a
shock caused by a large spatial velocity of \rat , but  the possibility that
magnetic energy contributes to power the observed diffuse emission has also been proposed \cite{rea09b}.

X-ray  upper limits at a level much smaller than the flux of \rat\ have been reported for other two RRATs \cite{kap09c}.
If the X-ray emission from RRATs is simply due to cooling, this could be caused by a larger
age of these two objects (their spin-down ages are 0.4 and 0.8 Myr,
compared to  $\tau_{SD}$=0.1 Myr   for \rat\ ). However, the significant uncertainties on distance
and absorption prevent a firm conclusion.

%%%%%%%%%%%%%%%%%%%%%%%%%%%%%%%%%%%%%%%%%%%%%%%%%%%%%%%%%%%%%%55
\begin{table}
\caption{Rotation Powered Pulsars with B$>$4 10$^{13}$ G.}
\label{tab:highB}
\begin{tabular}{lccccl}
\hline
\noalign{\smallskip}
Name  & Period  & $\pdot$       & B   & L$_x$  &   \\
      &  (s)  &  (s s$^{-1}$)  & (G) &  (erg s$^{-1}$) &     \\
\noalign{\smallskip}\svhline\noalign{\smallskip}
 J1847--0130~~ &  6.7   & 1.3 10$^{-12}$ & ~~9.4 10$^{13}$~~ & $<$5 10$^{33}$ &   \\[3pt]
 J1718--3718~~ &  3.4   & 1.6 10$^{-12}$ & ~~7.4 10$^{13}$~~ &   6 10$^{33}$ &   \\[3pt]
 J1814--1744~~ &  4.0   & 7.5 10$^{-13}$ & ~~5.5 10$^{13}$~~ &  $<$2 10$^{33}$&   \\[3pt]
 J1734--3333~~ &  1.2   & 2.3 10$^{-12}$ & ~~5.2 10$^{13}$~~ &  --  &   \\[3pt]
 J1846--0258~~ &  0.3   & 7.1 10$^{-12}$ & ~~4.9 10$^{13}$~~ &  4 10$^{34}$& in SNR Kes 75  \\[3pt]
 J1119--6127~~ &  0.4   & 4.0 10$^{-12}$ & ~~4.1 10$^{13}$~~ &  3 10$^{33}$ & in SNR G292.2--0.5  \\[3pt]
\hline
\noalign{\smallskip}\hline\noalign{\smallskip}
\end{tabular}
\end{table}
%%%%%%%%%%%%%%%%%%%%%%%%%%%%%%%%%%%%%%%%%%%%%%%%%%%%%%%%%%%%%%%%%%%%%%%%%%%%%%%%%%%%%%%%%%%

\section{Rotation-powered pulsars with high magnetic field}
\label{sec:rpp}

Table \ref{tab:highB} lists all the RPP for which the timing parameters
indicate a dipolar field
greater than,  or close to, the quantum critical value B$_c$ = $\frac{m^2 c^3}{\hbar
e}$=4.4 10$^{13}$ G.  Most of them
have spin periods of a few seconds and in the P-$\pdot$ diagram
lie in   the same region   of the AXPs/SGRs.
However, these four pulsars do not show magnetar characteristics,
such as the production of bursts/flares  or particularly strong and variable  emission.
In fact their X-ray luminosity is smaller than L$_{SD}$ \cite{piv00},
as expected for standard RPP.

The situation is quite different for the young PSR J1846--0258, located in the
supernova remnant Kes 75.
In this 0.3 s pulsar, the high inferred B derives from its large $\pdot$,  the highest of all RPPs.
Its spin-down luminosity of 8 10$^{36}$ erg s$^{-1}$ is certainly sufficient  to
power the  X-ray  emission  observed from the pulsar, L$_X$= 2.6 10$^{34}$ (d/6 kpc)$^2$ erg s$^{-1}$,
and  from the associated pulsar wind nebula,
L$_{PWN}$= 1.4 10$^{35}$ (d/6 kpc)$^2$ erg s$^{-1}$ \cite{ng08}.
Thus the discovery of   magnetar-like activity
from PSR J1846--0258 was quite surprising, because this pulsars was considered a
\textit{bona-fide} RPP, despite the lack of a radio detection
(easily explained as due to  an unfavorable orientation).
Four  short bursts from  PSR J1846--0258 were seen with RXTE  on May 31, 2006 \cite{gav08b}.
The onset of  the bursting activity  was coincident with a large
spin-up glitch and with the beginning of an enhancement
in the pulsed X-ray flux which lasted about one  month \cite{liv10}.
Spectral and flux variations associated with this event  were seen also
in the hard X-ray range \cite{kui09a}.

PSR J1119--6127  \cite{gon05,saf08}, has timing parameters very similar to those of PSR J1846--0258, and, like the latter,
it is located in a young SNR (G 292.2--0.5) and is surrounded by a pulsar wind nebula.
Its X-ray emission shows a large pulsed fraction and a spectrum consisting of a thermal-like
component with temperature kT$_{BB}\sim$0.2 keV and a hard power-law.
The strong thermal emission, quite remarkable for a young pulsar, might be the result of some recent magnetar-like activity, similar to that observed in PSR J1846--0258.
It would be important to monitor this pulsar, and other RPP with similar
parameters, to look for bursts.

\begin{figure}[b]
%\sidecaption
\includegraphics[scale=.4, angle=0]{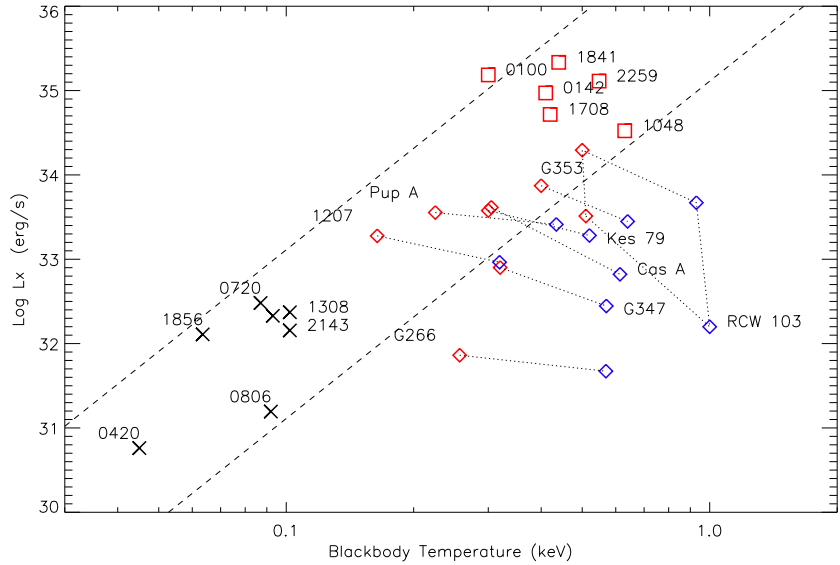}
\caption{Temperature-Luminosity diagram for different classes of INS
(XDINS ($\times$), CCO ($\diamond$), persistent AXP (squares)).
For the sake of uniformity, all the temperatures are  from blackbody fits
(model atmospheres generally yield values smaller by a factor $\sim$2--3).
For the CCO the luminosities and temperatures of the two-blackbody fits are indicated.
For RCW 103 the figure shows two intensity states.
In AXPs the contribution to the luminosity from the power-law
spectral components are neglected.
The lines indicate the blackbody relations for emitting radii of 1 km and 10 km.}
\label{fig:HR}
\end{figure}

\section{Conclusions}

X-ray observations have revealed  an unexpected variety of manifestations of INS.
The thermal components in these objects span more than four orders of magnitude in luminosity (see Fig. \ref{fig:HR}), with a dependence on temperature broadly consistent with the blackbody relation L$\propto$T$^4$ for objects of similar size. The observed spectral and timing properties indicate that different energy sources, besides the obvious cooling of the primeval internal heat, are
at play in powering such thermal-like components.

The effects  of a strong and dynamic magnetic field are especially evident in the paroxysmic
behavior of SGRs and AXPs, but it is also possible that the variability
observed in the XDIN \xdzerosette\ be due to changes in the magnetic field configuration, and
certainly the magnetic field plays an important role in the thermal
evolution of INS. This is well demonstrated by the relatively high surface temperatures of
XDINS.  It is now clear that at least a fraction of the enigmatic CCOs are   young
pulsars with very weak magnetic fields.  However, also in these sources, localized regions of
stronger field might be present and be responsible for the small emitting areas inferred
from the spectral fits.

The magnetar-like behavior of  PSR J1846--0258,
together with the recent discovery of PSR J1622--4950 \cite{lev10},
the first ''radio-selected'' magnetar,  point to a closer  connection between
radio pulsars and magnetars than previously thought.
It is also noteworthy that SGR 0418+5729, one of the most recently
discovered SGRs, has an inferred (dipolar) field smaller than $\sim$3 10$^{13}$ G.

The data collected in the latest years, besides enlarging the sample of INS,
are leading to recognize objects with ''intermediate'' properties.
While this makes the boundaries between the different INS classes  less
well defined, it gives the prospects of unifying in a global coherent picture
all the varieties  of  INS manifestations.

%

%%%%%%%%%%%%%%%%%%%%%%%% referenc.tex %%%%%%%%%%%%%%%%%%%%%%%%%%%%%%
% sample references
% %
% Use this file as a template for your own input.
%
%%%%%%%%%%%%%%%%%%%%%%%% Springer-Verlag %%%%%%%%%%%%%%%%%%%%%%%%%%
%
% BibTeX users please use
% \bibliographystyle{}
% \bibliography{}
%
%\bibliographystyle{plain}
%%%%%\bibliography{axpsgr3}

\end{document}